\documentclass[pra,superscriptaddress,twocolumn]{revtex4-2}

\usepackage{enumerate}
\usepackage{amsfonts,amssymb,amsmath}
\usepackage[]{graphics,graphicx,epsfig}
\usepackage{amsthm}
\usepackage{graphicx}
\usepackage{dcolumn}
\usepackage{natbib}
\usepackage{color}
\usepackage{multirow}
\usepackage{ulem}
\usepackage{bm}
\usepackage{url}
\usepackage{float}
\usepackage{ragged2e}

\usepackage{tikz}
\usepackage{environ}
\usetikzlibrary{automata,arrows,positioning,calc,shapes}
\usepackage[labelformat=simple]{subcaption}
\usepackage[justification=justified,format=plain]{caption}

\usepackage[colorlinks = true, citecolor=blue, linkcolor=blue, urlcolor=blue]{hyperref}
\newcommand*{\fullref}[1]{\hyperref[{#1}]{\autoref*{#1} \nameref*{#1}}}

\newcommand{\tr}[1]{\text{Tr}\left[#1\right]}

\begin{document}


\title{Communication Cost in Simulating Unknown Entangled States}
\author{Kelvin Onggadinata}
\affiliation{Centre for Quantum Technologies,
National University of Singapore, 3 Science Drive 2, 117543 Singapore,
Singapore}
\affiliation{Department of Physics,
National University of Singapore, 3 Science Drive 2, 117543 Singapore,
Singapore}

\author{Pawe{\l} Kurzy{\'n}ski}
\email{pawel.kurzynski@amu.edu.pl}
\affiliation{ Institute of Spintronics and Quantum Information, Faculty of Physics, Adam Mickiewicz University, Uniwersytetu Pozna{\'n}skiego 2, 61-614 Pozna\'n, Poland}
\affiliation{Centre for Quantum Technologies,
National University of Singapore, 3 Science Drive 2, 117543 Singapore,
Singapore}

\author{Dagomir Kaszlikowski}
\email{phykd@nus.edu.sg}
\affiliation{Centre for Quantum Technologies,
National University of Singapore, 3 Science Drive 2, 117543 Singapore,
Singapore}
\affiliation{Department of Physics,
National University of Singapore, 3 Science Drive 2, 117543 Singapore,
Singapore}

\date{\today}


\begin{abstract}
We demonstrate how to universally simulate ensemble statistics of projective local measurements on any $n$-qubit state shared among $n$ observers with classical communication and shared randomness. Our technique originates from protocols designed to simulate quantum non-locality [in {\it Horizons of the Mind}, Springer, Cham (2014)] and classical simulation of quantum circuits [Phys. Rev. Lett. 115, 070501 (2015)]. The protocol preserves three crucial aspects of the simulated quantum scenario in contrast to other approaches: no involvement of additional parties, none of the observers knows the global state of the system, and local measurement settings remain undisclosed.

\end{abstract}

\maketitle


\section{Introduction}

Classical simulation of local measurements on entangled quantum states is a fundamental problem that is particularly interesting in quantum information theory, since it relates quantum resources to classical ones. It was shown that such measurements can be classically simulated using a finite amount of communication if the prior information shared by the parties is infinite \cite{Masar}. Toner and Bacon demonstrated that if Alice and Bob share two real-valued random variables $\hat{\lambda}_1,
\hat{\lambda}_2 \in \mathbb{R}^3$, represented as unit vectors, they can simulate arbitrary local projective measurements on maximally entangled two-qubit state with a single bit of communication \cite{TB}. This result was later generalised to POVM \cite{DLR}. A single bit of communication is also enough to simulate local projective measurements on weakly entangled two-qubit states \cite{RQ} and there is a strong numerical evidence that one bit of communication is also enough to simulate measurements on partially entangled pure two-qubit states \cite{Valerio}. However, simulation of general measurements on an arbitrary two-qubit state requires two bits of communication \cite{RTQ}. Simulations of local measurements on two higher-dimensional systems \cite{BrassardN,DLR2,Vertesi} and multi-qubit states \cite{Branciard,Branciard2,BrassardMulti,Brassard2} were also considered. 

In our paper we handle this problem in a different way, focusing on classical simulation of quantum theory in phase space where quasi-probability distributions are a necessity. As a main tool we use the protocol to simulate quasi-probabilities with positive probability distributions first proposed by Abramsky and Brandenburger \cite{abramsky2014operational} and later by Pashayan {\it et al.} \cite{pashayan2015estimating} (these two protocols are essentially the same). For more details see also our complementary work \cite{us}. While our protocol requires an average communication cost that scales exponentially with the number of qubits, it preserves crucial features of the simulated quantum scenario. First, in the quantum scenario $n$ parties share an $n$-qubit state $\rho$ produced by some source. Without a prior knowledge of the source, the parties know nothing about the state of the system. In addition, each party locally chooses a measurement setting and does not reveal it to the others. Finally, no one else is involved in the scenario. We speculate that the exponential cost of communication is the price one has to pay to retain these features.

The structure of the paper is as follows. We first introduce a quasi-probabilistic representation of $n$-qubit states. In this representation states correspond to classical probability distributions, but measurements correspond to quasi-stochastic processes. Next we show how these quasi-stochastic processes can be classically simulated with standard stochastic processes. We discuss this simulation in the setting in which observers are spatially separated and highlight the necessity of communication. This leads to our protocol for a universal classical simulation of local projective measurements on $n$-qubits. 

\section{Quasi-stochastic representation of qubits}

We demonstrate how to map $n$-qubit states to positive probability distributions and local projective measurements as quasi-stochastic processes (we can also simulate POVMs but this is beyond the scope of this paper). Such representations have been known since Wigner's original paper \cite{wigner1932quantum} and subsequently formalised via so-called {\it frames} \cite{ferrie2008frame, ferrie2009framed}. However, here we chose a specific implementation of frames, most suitable to our approach. 

To begin, we observe that any positive probability distribution over $M$ bits, i.e., bit strings $a_1a_2\dots a_M$, $a_i=\pm 1$, can be always written as follows
\begin{eqnarray}
   p(a_1a_2\dots a_M) &=& \nonumber\\
   \frac{1}{2^M}(1+\sum_{i=1}^Ma_i\langle A_i\rangle &+&
   \sum_{\substack{i,j=1\\i<j}}^M a_ia_j\langle A_iA_j\rangle + \dots  \nonumber \\
   &+& a_1a_2\dots a_M\langle A_1A_2\dots A_M\rangle), 
\end{eqnarray}
where the modulus of each correlation function $\langle A_i\rangle, \langle A_iA_j\rangle,\dots, \langle A_1A_2\dots A_M\rangle$ is not greater than one. Of course, this does not guarantee positivity, which requires rather complicated constraints on the correlation functions. We can now represent $n$-qubit quantum states using such $M$-bit probability distributions, where $M=2n$.  

As a starting point, let us first represent a single qubit with a Bloch-vector ${\bf{s}}=(s_x,s_y,s_z)$ as
\begin{equation}
    p(aa') = \frac{1}{4}(1+\frac{1}{\sqrt{3}}(s_x a + s_y a' +s_z aa')).
\end{equation}
following the representation in \cite{onggadinata2023qubits}. To complete the picture we now need to represent projective measurements along some arbitrary direction ${\bf{m}}$ on the Bloch sphere. Since the state is described by a positive probability distribution, the measurement is a quasi-stochastic process converting a bit string $a a'$ to a single bit $\alpha=\pm 1$:
\begin{equation}
    \eta_{{\bf{m}}}(\alpha|aa')=\frac{1}{2}(1+\sqrt{3}\alpha(m_x a +m_y a' + m_z aa')).
\end{equation}
It is indeed quasi-stochastic because some transition probabilities are negative. These measurement probabilities follow the Born rule:
\begin{equation}
    q(\alpha|{\bf{m}})=\sum_{a,a'}\eta_{{\bf{m}}}(\alpha|aa')p(aa')=\frac{1}{2}(1+\alpha {\bf{s}}\cdot{\bf{m}}).
\end{equation}
It is then evident that this formalism, i.e., a positive probability distribution $p(aa')$ and the quasi-stochastic process $\eta_{{\bf{m}}}(\alpha|aa')$ are equivalent to quantum mechanics of a single qubit in the Heisenberg picture. 

We can simplify the above formulas by introducing a set of vectors ${\bf{n}}_{aa'}=\frac{1}{\sqrt{3}}(a,a',aa')$. It is straightforward to check that they form a three-dimensional tetrahedron and 
\begin{eqnarray}
&&p(aa') = \frac{1}{4}(1+{\bf{s}}\cdot{\bf{n}}_{aa'}) \nonumber\\
&&\eta_{{\bf{m}}}(\alpha|aa')=\frac{1}{2}(1+3\alpha{\bf{m}}\cdot{\bf{n}}_{aa'}).
\end{eqnarray}
    
Before we generalize this formalism to $n$-qubits let us consider another special case: two qubits. A two-qubit state $\rho$ is described by two local Bloch vectors, ${\bf{s}}$ and ${\bf{r}}$, and a $3\times 3$ correlation matrix ${\bf{T}}$
\begin{equation}
    \rho = \frac{1}{4} (1+{\bf{s}}\cdot{\boldsymbol{\sigma}}\otimes I +I\otimes {\bf{r}}\cdot{\boldsymbol{\sigma}}+\sum_{k,l}T_{kl}\sigma_k\otimes\sigma_l),
\end{equation}
where ${\boldsymbol{\sigma}}=(\sigma_x,\sigma_y,\sigma_z)$ and $\sigma_i$ ($i=x,y,z$) are Pauli matrices. This state is mapped to a four-bit positive probability distribution 
\begin{eqnarray}
  &&p(aa',bb')=\frac{1}{16}(1+{\bf{s}}\cdot{\bf{n}}_{aa'}+{\bf{r}}\cdot{\bf{n}}_{bb'}+{\bf{n}}_{aa'}\cdot {\bf{T}}{\bf{n}}_{bb'}).
\end{eqnarray}
Local measurements are now represented by the same quasi-stochastic processes as in the single qubit case, i.e.,
\begin{equation}
    q(\alpha,\beta|{\bf{m}}_A,{\bf{m}}_B) = \sum_{a,a',b,b'}\eta_{{\bf{m}}_A}(\alpha|aa')\eta_{{\bf{m}}_B}(\beta|bb')p(aa',bb'),
\end{equation}
where the probability distribution $q$ is what quantum theory predicts.
If one wishes to perform a global two-qubit measurement, they have to use an appropriate quasi-stochastic process 
\begin{equation}
\eta(\alpha\beta|aa',bb') = c_{aa'bb'\alpha\beta}+{\bf{s}}_{\alpha\beta}\cdot {\bf{n}}_{aa'}\otimes {\bf{n}}_{bb'},
\end{equation}
where $\sum_{\alpha\beta}c_{aa'bb'\alpha\beta}=1$ and ${\bf{s}}_{\alpha\beta}$ is a nine-dimensional real vector. This, however, is beyond the scope of this paper (and it can be easily found). 

Continuing the trend, an arbitrary $n$-qubit probability distribution reads
\begin{eqnarray}
& &p(a_1 a'_1,a_2 a'_2,\dots,a_n a'_n) = \nonumber \\
& &\frac{1}{4^n}\left(1+\sum_{k=1}^n \chi_k(a_1 a'_1,a_2 a'_2,\dots,a_n a'_n)\right),
\end{eqnarray}
where
\begin{eqnarray}
& &\chi_k(a_1 a'_1,a_2 a'_2,\dots,a_n a'_n) = \nonumber \\
& &\quad \sum_{\kappa_k \in K_k}\left(\sum_{j_1\ldots j_k=x,y,z} T^{(\kappa_k)}_{j_1\ldots j_k}\prod_{i=1}^k({\mathbf{n}}_{a_i a'_i})_{j_i}\right),
\end{eqnarray}
and $K_k$ is the family of all $k$-qubit subsets $\kappa_k$. The element $T^{(\kappa_k)}_{j_1\ldots j_k}$ is the element of the correlation tensor for the $k$-qubit subset $\kappa_k$. In particular, for single-qubit subsets $\kappa_1$, the element $T^{(\kappa_1)}_{j_1}$ is the Bloch's vector coordinate. Finally, note that the number of subsets in $K_k$ is $\binom{n}{k}$. The non-negativity is guaranteed by the non-negativity of a represented quantum state $\rho$ via the formula
\begin{eqnarray}
& & p(a_1a_1',a_2a_2',\dots,a_na_n') \nonumber \\
& &\qquad \quad  = \tr{\rho \Pi_{a_1a_1'}\otimes \Pi_{a_2a_2'}\otimes\dots\otimes \Pi_{a_n a_n'}}, 
\end{eqnarray}
where $\Pi_{a_ia_i'}=\frac{1}{4}(1+{\bf {n}}_{a_ia_i'}\cdot\boldsymbol{\sigma})$ is an element of the symmetric, informationally complete (SIC)-POVM \cite{renes2004symmetric}. Finally, projective measurements on such $n$-qubit distributions are as discussed before.

For instance, the probability distribution for the three-qubit GHZ state $\frac{1}{\sqrt{2}}(|000\rangle+|111\rangle)$ reads
\begin{eqnarray}
    &&p_{GHZ}(aa',bb',cc')= \frac{1}{64} \times \nonumber\\
    && \Big[ 1+({\bf n}_{aa'})_z({\bf n}_{bb'})_z+({\bf n}_{aa'})_z({\bf n}_{cc'})_z+({\bf n}_{bb'})_z({\bf n}_{cc'})_z+ \nonumber \\
    && ({\bf n}_{aa'})_x({\bf n}_{bb'})_x({\bf n}_{cc'})_x -({\bf n}_{aa'})_x({\bf n}_{bb'})_y({\bf n}_{cc'})_y -  \nonumber \\
    &&  ({\bf n}_{aa'})_y({\bf n}_{bb'})_x({\bf n}_{cc'})_y - ({\bf n}_{aa'})_y({\bf n}_{bb'})_y({\bf n}_{cc'})_x \Big], 
\end{eqnarray}
which is equivalent to
\begin{eqnarray}
&& p_{GHZ}(aa',bb',cc') \nonumber\\
&& \quad = \frac{1}{16}\Big[1 + \frac{1}{3}(aa'bb'+aa'cc'+bb'cc') \nonumber \\ 
&& \qquad  + \frac{1}{3\sqrt{3}}(abc-ab'c'-a'bc'-a'b'c)\Big].
\end{eqnarray}

The presented picture offers a consistent representation of quantum theory where quantum states are equivalent to non-negative probability distributions and measurements are quasi-stochastic processes. Thus, quantum behaviour in our picture, is solely the property of measurements because we do not have quasi-probabilities in classical theory. 


\section{Classical simulation of quasi-stochastic processes}

Any quasi-stochastic process is described by a quasi-stochastic matrix, i.e., a matrix with entries that can be negative, but its columns sum up to one. In our case a quasi-stochastic process $\eta_{{\bf{m}}}(\alpha|aa')$, describing a quantum measurement, is a $2\times 4$ matrix. Such a matrix can always be written as a linear combination of two non-negative stochastic matrices $S^{\pm}$ weighted by a negative binary distribution (so-called {\it nebit}, see \cite{kaszlikowski2021little}):
\begin{equation}
    S = (1+\delta)S^{+} - \delta S^-
\end{equation}
with $\delta\geq 0$. We showed in \cite{onggadinata2023simulations} how to find a nebit decomposition of any quasi-stochastic matrix using the following components:
\begin{eqnarray}
\delta &=& \max\{0, -2\min_{k,l}S_{kl}\}\\
S^- &=& \frac{1}{2}\mathbf{1}\\
S^+ &=& \frac{1}{1+\delta}(S+\delta S^-)
\end{eqnarray}
where $\mathbf{1}$ is a $2\times 4$ matrix with all entries equal to $1$. Note that $1 \geq \delta \geq (\sqrt{3}-1)/2 \approx 0.366$, where the lower bound corresponds to measurements along direction ${\bf{m}}={\bf{x}},{\bf{y}},{\bf{z}}$ and the upper bound corresponds to measurements along tetrahedral directions ${\bf{m}}={\bf{n}}_{aa'}$.

We now show how to simulate an action of a quasi-stochastic process $S$, representing a projective measurement given by $\eta_{{\bf{m}}}(\alpha|aa')$, on a non-negative single-qubit distribution $p(aa')$. This is a classical simulation that uses a standard probability calculus with non-negative probabilities only. It was first discovered, although in a different but related scenario, by Abramsky and Brandenburger \cite{abramsky2014operational} and later by Pashayan {\it et. al.} \cite{pashayan2015estimating}. Additional details can be found in our complementary work \cite{us}.

To simulate $S$, we use a binary random bit $r=\pm 1$, distributed with probability $\{t,1-t\}$
\begin{equation}
    t = \frac{1+\delta}{1+2\delta}\, .
\end{equation}
Note that $t>1-t$ since $1+\delta > \delta$. We then perform a controlled operation conditional on the outcome: apply $S^+$ if $r=+1$, else apply $S^-$. We get
\begin{eqnarray}
q^{\pm}(\alpha) = \sum_{aa'} S^{\pm}(\alpha|aa')p(aa')
\end{eqnarray} 
Therefore, in a single round a string $aa'$ transforms into a single bit $\alpha$. However, we keep the track of $r$, hence we have a string $r\alpha$. 

After $N$ rounds we hold a table of such pairs. The next step is to modify it in the following way. For each event $\{r=-1,\alpha\}\equiv \{-1,\alpha\}$ we find an event $\{+1,\alpha\}$ and remove both of them from the table. The schematic representation of the removal procedure is shown below.
\begin{center} 
\begin{tabular}{ |c|| c | c | } 
 \hline
  & r & $\alpha$ \\ 
  \hline
  \hline
   1 & +1 & +1  \\ 
  \hline
  2 & +1 & -1 \\
  \hline
  3 & -1 & +1  \\
  \hline
 \vdots & \vdots & \vdots   \\
  \hline
  N & +1 & +1  \\
  \hline
\end{tabular} ~~ $\rightarrow$ ~~ 
\begin{tabular}{ | c | c | c | } 
 \hline
  & r & $\alpha$ \\ 
  \hline
  \hline
 {\color{red} 1} &  {\color{red} +1} &  {\color{red}  +1}  \\ 
  \hline
 2 & +1 & -1  \\
   \hline
 {\color{red}  3} & {\color{red}  -1} &  {\color{red} +1}  \\
  \hline
 \vdots & \vdots & \vdots   \\
    \hline
 N & +1 & +1  \\
  \hline
\end{tabular} ~~ $\rightarrow$ ~~ 
\begin{tabular}{ | c | c | c | } 
 \hline
 &  r & $\alpha$ \\ 
  \hline
   \hline
  1 & +1 & -1  \\
  \hline
  \vdots & \vdots & \vdots   \\
  \hline
  M & +1 & +1  \\
  \hline
\end{tabular} 
\end{center}
In the above $M$ is the total number of events that remained after the removal procedure. The end game is to get rid of all events $\{-1,\alpha\}$ from the table. If succeeded, we use the final table to evaluate the probabilities $q(\alpha)$. If failed, i.e., there are still some events  $\{-1,\alpha\}$ in the table for which there is no pair $\{+1,\alpha\}$, we terminate the simulation. However, for a sufficiently large number of rounds $N$, the number of events $\{+1,\alpha\}$ in the initial table is $N^+_{\alpha} \approx N t q^+(\alpha)$ and the number of events with $\{-1,\alpha\}$ is $N^-_{\alpha}\approx N (1-t) q^-(\alpha)$. After the removal the total number of remaining events $\{+1,\alpha\}$ is
\begin{eqnarray}
M_{\alpha} &=& (N^+_{\alpha}-N^-_{\alpha}) \approx N \left(tq^+(\alpha) - (1-t)q^-(\alpha)\right)=  \nonumber \\
& =& \frac{N}{1+2\delta}\left( (1+\delta)q^+(\alpha) - \delta q^-(\alpha) \right) = \nonumber \\
&=& \frac{N}{1+2\delta}\sum_{aa'}\eta_{\vec{m}}(\alpha|aa')p(aa').
\end{eqnarray}
In addition, the total number of all remaining events is 
\begin{eqnarray}
M&=&\sum_{\alpha} M_{\alpha} \approx \frac{N}{1+2\delta},
\end{eqnarray}
therefore 
\begin{equation}
\frac{M_{\alpha}}{M} \approx \sum_{aa'}\eta_{\vec{m}}(\alpha|aa')p(aa')=q(\alpha).     
\end{equation}
This confirms that our method simulates quasi-stochastic processes considered here.

In order to determine how the simulation's precision scales with the number of rounds $N$, one can use the Hoeffding's inequality \cite{pashayan2015estimating}. In particular, our estimate $M_{\alpha}/M$ is within $\epsilon$ of $q(\alpha)$ with a probability $1-\Delta$ if 
\begin{equation}
    M \geq \frac{1}{2\epsilon^2}\log \frac{2}{\Delta}\, ,
\end{equation}
hence 
\begin{equation}
    N \geq \frac{1+2\delta}{2\epsilon^2}\log \frac{2}{\Delta}\, .
\end{equation}


\section{Classical simulation of local measurement statistics on n-qubit state}

Let us frame our simulation scenario as a game. There are $n$ parties, named $A_i$ ($i=1,\ldots,n$), who are challenged to simulate statistics of local projective measurements on an unknown $n$-qubit state represented by $2n$-bit shared non-negative probability distribution. Each party randomly chooses their measurement settings. The parties are expected to use this shared randomness, their local measurement choices, and classical communication to generate a table of outcomes that mimics the one generated in a scenario in which they performed projective measurements on an actual unknown $n$-qubit state. 

The unknown $n$-qubit state is represented by a probability distribution $p(a_1a'_1,a_2a'_2,\ldots,a_na'_n)$. Therefore, a single copy of a system corresponds to strings $a_i a'_i$ shared between $A_i$. We assume that parties have access to $N$ copies of the system, hence $A_i$ has a list of $N$ $a_i a'_i$ bit strings. These lists can be represented as a table, like this one
\begin{center}
\begin{tabular}{ |c||c|c|c|c|c| } 
\hline
~ & $a_1$ & $a'_1$ & \ldots & $a_n$ & $a'_n$ \\
\hline
\hline
1 & +1 & -1 & \ldots & +1 & +1 \\ 
\hline
2 & -1 & -1 & \ldots & +1 & -1 \\ 
\hline
3 & -1 & +1 & \ldots & -1 & -1 \\ 
\hline
\vdots & \vdots & \vdots & \ldots & \vdots & \vdots \\ 
\hline
N & +1 & +1 & \ldots & -1 & +1 \\ 
\hline
\end{tabular}
\end{center}
We stress that $A_i$ has access only to two columns corresponding to $a_i$ and $a'_i$. This table represents classical statistics. The goal is to transform it into one that mimics local measurement outcomes on an actual $n$-qubit state.

\begin{figure}
\centering
\begin{subfigure}[t]{1\linewidth}
\centering
\resizebox{1\linewidth}{!}{%
\begin{tikzpicture}[
roundnode/.style={ellipse, draw=black, fill=white, very thick, scale=1,align=center, minimum width=2.5cm},
squarednode/.style={rectangle, draw=black, fill=white, very thick, scale=1,align=center, minimum width = 2.5cm, minimum height = 1.5cm},
]
\node[roundnode] (S) at (0,0) {\textbf{Source}\\Choose $p(\vec{a})$};
\node[squarednode] (A) at (-4,-4) {$\mathbf{A_1}$\\Choose $\mathbf{m}_1$\\Calc. $r_1,\alpha_1$};
\node[squarednode] (B) at (0,-4) {$\mathbf{A_2}$\\Choose $\mathbf{m}_2$\\Calc. $r_2,\alpha_2$};
\node at (2,-4) {\Large{$\dots$}};
\node[squarednode] (Z) at (4,-4) {$\mathbf{A_n}$\\Choose $\mathbf{m}_n$\\Calc. $r_n,\alpha_n$};
\path[>=stealth, ->, draw, thick]
(S) edge[above left] node{$a_1,a_1'$} (A.north)
(S) edge[right] node{$a_2,a_2'$} (B.north)
(S) edge[above right] node{$a_n,a_n'$} (Z.north)
(B) edge[above] node{$r_2,\alpha_2$} (A)
(Z) edge[bend left, above] node{$r_n,\alpha_n$} (A);
\end{tikzpicture}
}
\end{subfigure}
\begin{subfigure}[t]{1\linewidth}
\centering
\resizebox{1\linewidth}{!}{%
\begin{tikzpicture}[
roundnode/.style={ellipse, draw=black, fill=white, very thick, scale=1,align=center, minimum width=2.5cm},
squarednode/.style={rectangle, draw=black, fill=white, very thick, scale=1,align=center, minimum width = 2.5cm, minimum height = 1.5cm},
]
\node[squarednode] (A) at (-4,0) {$\mathbf{A_1}$\\Post-process};
\node[squarednode] (B) at (0,0) {$\mathbf{A_2}$};
\node at (2,0) {\Large{$\dots$}};
\node[squarednode] (Z) at (4,0) {$\mathbf{A_n}$};
\path[>=stealth, ->, draw, thick]
(A.east) edge[above left,pos=0.6] node{$\alpha_2$} (B)
(A.north east) edge[bend left, above] node{$\alpha_n$} (Z);
\draw[>=stealth, double, ->, thick] (A.south) to ([shift={(0cm,-1cm)}]A.south);
\draw[>=stealth, double, ->, thick] (B.south) to ([shift={(0cm,-1cm)}]B.south);
\draw[>=stealth, double, ->, thick] (Z.south) to ([shift={(0cm,-1cm)}]Z.south);
\node at ([shift={(0cm,-1.2cm)}]A.south) {$\alpha_1$};
\node at ([shift={(0cm,-1.2cm)}]B.south) {$\alpha_2$};
\node at ([shift={(0cm,-1.2cm)}]Z.south) {$\alpha_n$};
\end{tikzpicture}
}
\end{subfigure}
\caption{\justifying{Our proposed protocol to locally simulate ensemble statistics of any $n$-partite qubit states using shared randomness and classical communication. (Top) The first part of the protocol involved a common source who send bits $a_ka_k$ to the $k$-th observer as described by probability distribution $p(\vec{a})\equiv p(a_1a_1'a_2a_2'\dots a_na_n')$. Each party then chooses a measurement independently and determine their measurement results as $r_k,\alpha_k$. Every party then sends their outcome results to $A_1$ to perform post-processing. (Bottom) The updated outcomes after post-processing will then be send back in which they will announce the results.}}
\label{fig: our protocol}
\end{figure}
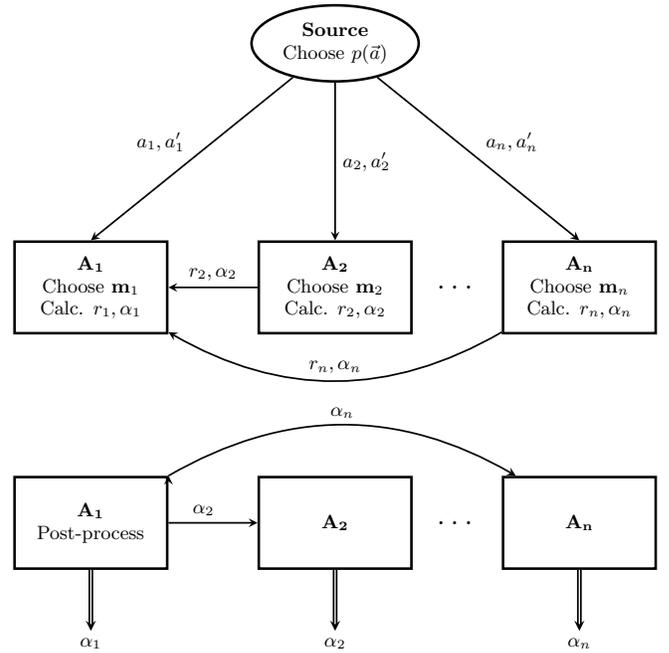

The simulation scenario is a straightforward extension of the ideas from the previous sections. The building blocks are non-negative probability distributions $p(a_1a'_1,a_2a'_2,\ldots,a_na'_n)$ and quasi-stochastic transformations $\eta_{{\bf m}_i}(\alpha_i|a_i a'_i)$, where ${\bf m}_i$ is the i-th qubit measurement direction. These quasi-stochastic transformations are simulated  with the help of local random bits $r_i=\pm 1$, distributed with probabilities $t_i=\frac{1+\delta_i}{1+2\delta_i}$ and $1-t_i=\frac{\delta_i}{1+2\delta_i}$, followed by the removal of certain detection events from the outcome list. Importantly, the removal procedure requires classical communication. Here is the breakdown of the protocol with Fig. \ref{fig: our protocol} illustrating the sequence of the protocol. 

\subsubsection*{Protocol}

Each party $A_i$ chooses a direction ${\bf m}_i$ of local measurement and finds the corresponding $\delta_i$ and the decomposition $\eta_{{\bf m}_i} = (1+\delta_i)S^+_i - \delta_i S^-_i$, where $S^{\pm}_i$ are standard stochastic transformations. Next, $A_i$ generates $N$ random bits $r_i$ and generates $\alpha_i$ by applying $S^{+}_i$ to $a_i$ and $a'_i$ (if $r_i=+1$) or $S^{-}_i$ (if $r_i=-1$). At this stage the original table generated by $B$ is locally transformed into a new one, say,
\begin{center}
\begin{tabular}{ |c||c|c|c|c|c| } 
\hline
~ & $r_1$ & $\alpha_1$ & \ldots & $r_n$ & $\alpha_n$ \\
\hline
\hline
1 & +1 & +1 & \ldots & -1 & +1 \\ 
\hline
2 & -1 & -1 & \ldots & -1 & -1 \\ 
\hline
3 & +1 & -1 & \ldots & +1 & -1 \\ 
\hline
\vdots & \vdots & \vdots & \ldots & \vdots & \vdots \\ 
\hline
N & +1 & +1 & \ldots & -1 & -1 \\ 
\hline
\end{tabular}
\end{center}

Note that $A_i$ cannot individually remove events from the list because they do not know the other parties' outcomes. Therefore $A_i$ ($i=2,\ldots,n$) sends their results to $A_1$. Communication of the column $\alpha_i$ requires $N$ bits, however communication of the column $r_i$ requires on average $N H(\delta_i)$ bits, where $H(\delta_i)=-t_i\log_2 t_i -(1-t_i)\log_2 (1-t_i)$ is the binary Shannon entropy. After receiving all the information $A_1$ calculates $r=r_1 r_2 \ldots r_N$, where $r=\pm 1$ indicates that the corresponding joint event occurs with {\it positive/negative} quasi-probability. Only now the removal procedure can be applied. The corresponding table changes to
\begin{center}
\begin{tabular}{ |c||c|c|c|c| } 
\hline
~ & $r$ & $\alpha_1$ & \ldots & $\alpha_n$ \\
\hline
\hline
1 & +1 & +1 & \ldots & +1 \\ 
\hline
{\color{red} 2} & {\color{red}+1} & {\color{red}-1} & {\color{red}\ldots} & {\color{red}-1} \\ 
\hline
{\color{red} 3} & {\color{red}-1} & {\color{red}-1 }& {\color{red}\ldots} & {\color{red}-1} \\ 
\hline
\vdots & \vdots & \vdots & \ldots & \vdots \\ 
\hline
N & +1 & +1 & \ldots & -1 \\ 
\hline
\end{tabular} ~~$\rightarrow$~~
\begin{tabular}{ |c||c|c|c|c| } 
\hline
~ & $r$ & $\alpha_1$ & \ldots & $\alpha_n$ \\
\hline
\hline
1 & +1 & +1 & \ldots & +1 \\ 
\hline
\vdots & \vdots & \vdots & \ldots & \vdots \\ 
\hline
M & +1 & +1 & \ldots & -1 \\ 
\hline
\end{tabular}
\end{center}
After the removal it contains M entries, where $M=fN$ and 
\begin{equation}
f=\prod_{i=1}^n \frac{1}{(1+2\delta_i)}
\end{equation}
is the fraction of remaining entries. Note that 
\begin{equation}\label{scaling}
  \left(\frac{1}{(1+2\delta_{max})}\right)^n \leq f \leq \left(\frac{1}{(1+2\delta_{min})}\right)^n,
\end{equation}
where $\delta_{max}=\max \{\delta_i\}_{i=1}^n$ and $\delta_{min}=\min \{\delta_i\}_{i=1}^n$.

In the last step of the protocol $A_1$ sends to $A_i$ the $\alpha_i$ column of the modified table. It requires $fN$ bits of communication. The communication part of the protocol is schematically presented in Fig. \ref{f1}. In total it requires on average
\begin{equation}
N\left(n-1+\sum_{i=2}^n H(\delta_i) +(n-1)f\right) 
\end{equation}
bits of communication to generate a table that mimics $M=fN$ rounds of local projective measurements on an unknown $n$-qubit state. This corresponds to
\begin{equation}\label{eq: communication cost per round}
\frac{1}{f}\left(n-1+\sum_{i=2}^n H(\delta_i) +(n-1)f\right)    
\end{equation}
bits of communication per one entry in the final table. Due to (\ref{scaling}), the above scales as $O(n \times e^n)$. This suggests that simulation cost of multipartite quantum nonlocality with classical communication protocols for an unknown state and with undisclosed settings scales exponentially with the number of measurements. It is an open question if this scaling can be improved.

\begin{figure}[!t]
    \begin{subfigure}[t]{1\linewidth}
    {\centering
    \begin{tikzpicture}[->, >=stealth', auto, semithick, node distance=3cm]
    \tikzstyle{every state}=[fill=white,draw=black,thick,text=black,scale=1]
    \node[state]    (A)                     {$A_1$};
    \node[state]    (B)[right of=A]   		{$A_2$};
    \path
    (A) edge[bend right, below] 	node{$f$}      (B)
    (B) edge[above] 	node{$1+H(\delta_2)$} 	(A);
    \end{tikzpicture}
    \caption*{Bipartite scenario 
    \vspace{5mm}}}
    \end{subfigure}
    \begin{subfigure}[t]{1\linewidth}
    {\centering
    \begin{tikzpicture}[->, >=stealth', auto, semithick, node distance=3.5cm]
    \tikzstyle{every state}=[fill=white,draw=black,thick,text=black,scale=1]
    \node[state]    (A)                     {$A_1$};
    \node[state]    (B)[left of=A]   		{$A_2$};
    \node[state]    (C)[above of=A]     {$A_3$};
    \node[state]    (D)[right of=A]     {$A_4$};
    \node[state]    (Z)[below of=A]     {$A_{n}$};
    \draw (290: 3.5cm) node[fill=white, text=black, scale=1, rotate=35]   (l1) {\Large{$\ldots$}};
    \draw (335: 3.5cm) node[fill=white, text=black, scale=1, rotate=70]   (l2) {\Large{$\ldots$}};
    \draw (315: 3.5cm) node[state] (E) {$A_i$};
    \path
    (A) edge[bend left, below]     node{$f$}    (B)
        edge[bend left, left]      node{$f$}    (C)
        edge[bend left, above]     node{$f$}    (D)
        edge[bend left, right,pos=0.8]     node{$f$}    (Z)
        edge[] (E)
    (E) edge[] (A)
    (B) edge[above] 	node{$1+H(\delta_2)$} 	 (A)
    (C) edge[right, pos=0.3]     node{$1+H(\delta_3)$}   (A)
    (D) edge[below]     node{$1+H(\delta_4)$} (A)
    (Z) edge[left,pos=0.3]     node{$1+H(\delta_{n})$} (A);
    \end{tikzpicture}
    \caption*{Multipartite scenario}}
    \end{subfigure}
\caption{\justifying{Illustration of the protocol, where the communication is given per one entry in the final table.}}
    \label{f1}
\end{figure}

\subsubsection*{Example}

Here, we show an example of communication cost of Mermin's $n$-partite Bell test \cite{mermin1990extreme}.  An initial GHZ state is shared amongst $n$ observers who measure local Pauli observables. As already mentioned, quasi-stochastic processes representing any of the Pauli measurements have $\delta = (\sqrt{3}-1)/2 \approx 0.366$. Thus, $H(\delta_i) = H(\delta)\approx 0.963$ for all $i=1,\dots, n$. Additionally, we have 
\begin{equation}
    f = \prod_{i=1}^n \frac{1}{1+2\delta_i} \approx \frac{1}{1.732^n}.
\end{equation}
From \eqref{eq: communication cost per round}, the total cost of communication per one entry in the final table is
\begin{equation}
    (1.732)^n(n-1)\left(1.963+\frac{1}{1.732^n}\right)
\end{equation}
bits. 


\section{Conclusions}

Using techniques developed in \cite{abramsky2014operational} and \cite{pashayan2015estimating} we showed how to simulate ensemble statistics of $n$ local projective measurements on any unknown $n$-qubit state. By unknown we mean that the global state is never revealed to the observers. They only have access to information about their local parts. Moreover, in our protocol local measurement settings stay private. This sets our protocol apart from previous ones in the literature on the topic. If the protocol cannot be further improved, it may explain why our communication scales exponentially with the number of observers. Whether this scaling is a consequence of not knowing the state remains an open question at this moment. 

Although we presented our simulation at the ensemble level, it also works as a \textit{single-shot} simulation, where a finite number of source samples is used to generate a single measurement outcome. This type of simulation yields noisy measurements, the extent of which diminishes with an increasing number of source samples. We have opted not to present the single-shot version to maintain the clarity of the paper.

A straightforward modification of the protocol leads to simulations of any $n$-partite $d$-dimensional quantum systems, which is fairly obvious as such systems can be simulated with quantum computer operating on many qubits.


\section*{Acknowledgements}

This research is supported by the National Research Foundation, Singapore, and A*STAR under the CQT Bridging Grant. PK is supported by the Polish National Science Centre (NCN) under the Maestro Grant no. DEC-2019/34/A/ST2/00081.



\bibliographystyle{apsrev4-2}
\bibliography{ref.bib}

\begin{thebibliography}{24}%
\makeatletter
\providecommand \@ifxundefined [1]{%
 \@ifx{#1\undefined}
}%
\providecommand \@ifnum [1]{%
 \ifnum #1\expandafter \@firstoftwo
 \else \expandafter \@secondoftwo
 \fi
}%
\providecommand \@ifx [1]{%
 \ifx #1\expandafter \@firstoftwo
 \else \expandafter \@secondoftwo
 \fi
}%
\providecommand \natexlab [1]{#1}%
\providecommand \enquote  [1]{``#1''}%
\providecommand \bibnamefont  [1]{#1}%
\providecommand \bibfnamefont [1]{#1}%
\providecommand \citenamefont [1]{#1}%
\providecommand \href@noop [0]{\@secondoftwo}%
\providecommand \href [0]{\begingroup \@sanitize@url \@href}%
\providecommand \@href[1]{\@@startlink{#1}\@@href}%
\providecommand \@@href[1]{\endgroup#1\@@endlink}%
\providecommand \@sanitize@url [0]{\catcode `\\12\catcode `\$12\catcode
  `\&12\catcode `\#12\catcode `\^12\catcode `\_12\catcode `\%12\relax}%
\providecommand \@@startlink[1]{}%
\providecommand \@@endlink[0]{}%
\providecommand \url  [0]{\begingroup\@sanitize@url \@url }%
\providecommand \@url [1]{\endgroup\@href {#1}{\urlprefix }}%
\providecommand \urlprefix  [0]{URL }%
\providecommand \Eprint [0]{\href }%
\providecommand \doibase [0]{https://doi.org/}%
\providecommand \selectlanguage [0]{\@gobble}%
\providecommand \bibinfo  [0]{\@secondoftwo}%
\providecommand \bibfield  [0]{\@secondoftwo}%
\providecommand \translation [1]{[#1]}%
\providecommand \BibitemOpen [0]{}%
\providecommand \bibitemStop [0]{}%
\providecommand \bibitemNoStop [0]{.\EOS\space}%
\providecommand \EOS [0]{\spacefactor3000\relax}%
\providecommand \BibitemShut  [1]{\csname bibitem#1\endcsname}%
\let\auto@bib@innerbib\@empty
\bibitem [{\citenamefont {Massar}\ \emph {et~al.}(2001)\citenamefont {Massar},
  \citenamefont {Bacon}, \citenamefont {Cerf},\ and\ \citenamefont
  {Cleve}}]{Masar}%
  \BibitemOpen
  \bibfield  {author} {\bibinfo {author} {\bibfnamefont {S.}~\bibnamefont
  {Massar}}, \bibinfo {author} {\bibfnamefont {D.}~\bibnamefont {Bacon}},
  \bibinfo {author} {\bibfnamefont {N.~J.}\ \bibnamefont {Cerf}},\ and\
  \bibinfo {author} {\bibfnamefont {R.}~\bibnamefont {Cleve}},\ }\href
  {https://doi.org/10.1103/PhysRevA.63.052305} {\bibfield  {journal} {\bibinfo
  {journal} {Phys. Rev. A}\ }\textbf {\bibinfo {volume} {63}},\ \bibinfo
  {pages} {052305} (\bibinfo {year} {2001})}\BibitemShut {NoStop}%
\bibitem [{\citenamefont {Toner}\ and\ \citenamefont {Bacon}(2003)}]{TB}%
  \BibitemOpen
  \bibfield  {author} {\bibinfo {author} {\bibfnamefont {B.~F.}\ \bibnamefont
  {Toner}}\ and\ \bibinfo {author} {\bibfnamefont {D.}~\bibnamefont {Bacon}},\
  }\href {https://doi.org/10.1103/PhysRevLett.91.187904} {\bibfield  {journal}
  {\bibinfo  {journal} {Phys. Rev. Lett.}\ }\textbf {\bibinfo {volume} {91}},\
  \bibinfo {pages} {187904} (\bibinfo {year} {2003})}\BibitemShut {NoStop}%
\bibitem [{\citenamefont {Degorre}\ \emph {et~al.}(2005)\citenamefont
  {Degorre}, \citenamefont {Laplante},\ and\ \citenamefont {Roland}}]{DLR}%
  \BibitemOpen
  \bibfield  {author} {\bibinfo {author} {\bibfnamefont {J.}~\bibnamefont
  {Degorre}}, \bibinfo {author} {\bibfnamefont {S.}~\bibnamefont {Laplante}},\
  and\ \bibinfo {author} {\bibfnamefont {J.}~\bibnamefont {Roland}},\ }\href
  {https://doi.org/10.1103/PhysRevA.72.062314} {\bibfield  {journal} {\bibinfo
  {journal} {Phys. Rev. A}\ }\textbf {\bibinfo {volume} {72}},\ \bibinfo
  {pages} {062314} (\bibinfo {year} {2005})}\BibitemShut {NoStop}%
\bibitem [{\citenamefont {Renner}\ and\ \citenamefont {Quintino}(2023)}]{RQ}%
  \BibitemOpen
  \bibfield  {author} {\bibinfo {author} {\bibfnamefont {M.~J.}\ \bibnamefont
  {Renner}}\ and\ \bibinfo {author} {\bibfnamefont {M.~T.}\ \bibnamefont
  {Quintino}},\ }\href {https://doi.org/10.22331/q-2023-10-24-1149} {\bibfield
  {journal} {\bibinfo  {journal} {{Quantum}}\ }\textbf {\bibinfo {volume}
  {7}},\ \bibinfo {pages} {1149} (\bibinfo {year} {2023})}\BibitemShut
  {NoStop}%
\bibitem [{\citenamefont {Sidajaya}\ \emph {et~al.}(2023)\citenamefont
  {Sidajaya}, \citenamefont {Lim}, \citenamefont {Yu},\ and\ \citenamefont
  {Scarani}}]{Valerio}%
  \BibitemOpen
  \bibfield  {author} {\bibinfo {author} {\bibfnamefont {P.}~\bibnamefont
  {Sidajaya}}, \bibinfo {author} {\bibfnamefont {A.~D.}\ \bibnamefont {Lim}},
  \bibinfo {author} {\bibfnamefont {B.}~\bibnamefont {Yu}},\ and\ \bibinfo
  {author} {\bibfnamefont {V.}~\bibnamefont {Scarani}},\ }\href
  {https://doi.org/10.22331/q-2023-10-24-1150} {\bibfield  {journal} {\bibinfo
  {journal} {{Quantum}}\ }\textbf {\bibinfo {volume} {7}},\ \bibinfo {pages}
  {1150} (\bibinfo {year} {2023})}\BibitemShut {NoStop}%
\bibitem [{\citenamefont {Renner}\ \emph {et~al.}(2023)\citenamefont {Renner},
  \citenamefont {Tavakoli},\ and\ \citenamefont {Quintino}}]{RTQ}%
  \BibitemOpen
  \bibfield  {author} {\bibinfo {author} {\bibfnamefont {M.~J.}\ \bibnamefont
  {Renner}}, \bibinfo {author} {\bibfnamefont {A.}~\bibnamefont {Tavakoli}},\
  and\ \bibinfo {author} {\bibfnamefont {M.~T.}\ \bibnamefont {Quintino}},\
  }\href {https://doi.org/10.1103/PhysRevLett.130.120801} {\bibfield  {journal}
  {\bibinfo  {journal} {Phys. Rev. Lett.}\ }\textbf {\bibinfo {volume} {130}},\
  \bibinfo {pages} {120801} (\bibinfo {year} {2023})}\BibitemShut {NoStop}%
\bibitem [{\citenamefont {Brassard}\ \emph {et~al.}(1999)\citenamefont
  {Brassard}, \citenamefont {Cleve},\ and\ \citenamefont {Tapp}}]{BrassardN}%
  \BibitemOpen
  \bibfield  {author} {\bibinfo {author} {\bibfnamefont {G.}~\bibnamefont
  {Brassard}}, \bibinfo {author} {\bibfnamefont {R.}~\bibnamefont {Cleve}},\
  and\ \bibinfo {author} {\bibfnamefont {A.}~\bibnamefont {Tapp}},\ }\href
  {https://doi.org/10.1103/PhysRevLett.83.1874} {\bibfield  {journal} {\bibinfo
   {journal} {Phys. Rev. Lett.}\ }\textbf {\bibinfo {volume} {83}},\ \bibinfo
  {pages} {1874} (\bibinfo {year} {1999})}\BibitemShut {NoStop}%
\bibitem [{\citenamefont {Degorre}\ \emph {et~al.}(2007)\citenamefont
  {Degorre}, \citenamefont {Laplante},\ and\ \citenamefont {Roland}}]{DLR2}%
  \BibitemOpen
  \bibfield  {author} {\bibinfo {author} {\bibfnamefont {J.}~\bibnamefont
  {Degorre}}, \bibinfo {author} {\bibfnamefont {S.}~\bibnamefont {Laplante}},\
  and\ \bibinfo {author} {\bibfnamefont {J.}~\bibnamefont {Roland}},\ }\href
  {https://doi.org/10.1103/PhysRevA.75.012309} {\bibfield  {journal} {\bibinfo
  {journal} {Phys. Rev. A}\ }\textbf {\bibinfo {volume} {75}},\ \bibinfo
  {pages} {012309} (\bibinfo {year} {2007})}\BibitemShut {NoStop}%
\bibitem [{\citenamefont {V\'ertesi}\ and\ \citenamefont
  {Bene}(2009)}]{Vertesi}%
  \BibitemOpen
  \bibfield  {author} {\bibinfo {author} {\bibfnamefont {T.}~\bibnamefont
  {V\'ertesi}}\ and\ \bibinfo {author} {\bibfnamefont {E.}~\bibnamefont
  {Bene}},\ }\href {https://doi.org/10.1103/PhysRevA.80.062316} {\bibfield
  {journal} {\bibinfo  {journal} {Phys. Rev. A}\ }\textbf {\bibinfo {volume}
  {80}},\ \bibinfo {pages} {062316} (\bibinfo {year} {2009})}\BibitemShut
  {NoStop}%
\bibitem [{\citenamefont {Branciard}\ and\ \citenamefont
  {Gisin}(2011)}]{Branciard}%
  \BibitemOpen
  \bibfield  {author} {\bibinfo {author} {\bibfnamefont {C.}~\bibnamefont
  {Branciard}}\ and\ \bibinfo {author} {\bibfnamefont {N.}~\bibnamefont
  {Gisin}},\ }\href {https://doi.org/10.1103/PhysRevLett.107.020401} {\bibfield
   {journal} {\bibinfo  {journal} {Phys. Rev. Lett.}\ }\textbf {\bibinfo
  {volume} {107}},\ \bibinfo {pages} {020401} (\bibinfo {year}
  {2011})}\BibitemShut {NoStop}%
\bibitem [{\citenamefont {Branciard}\ \emph {et~al.}(2012)\citenamefont
  {Branciard}, \citenamefont {Brunner}, \citenamefont {Buhrman}, \citenamefont
  {Cleve}, \citenamefont {Gisin}, \citenamefont {Portmann}, \citenamefont
  {Rosset},\ and\ \citenamefont {Szegedy}}]{Branciard2}%
  \BibitemOpen
  \bibfield  {author} {\bibinfo {author} {\bibfnamefont {C.}~\bibnamefont
  {Branciard}}, \bibinfo {author} {\bibfnamefont {N.}~\bibnamefont {Brunner}},
  \bibinfo {author} {\bibfnamefont {H.}~\bibnamefont {Buhrman}}, \bibinfo
  {author} {\bibfnamefont {R.}~\bibnamefont {Cleve}}, \bibinfo {author}
  {\bibfnamefont {N.}~\bibnamefont {Gisin}}, \bibinfo {author} {\bibfnamefont
  {S.}~\bibnamefont {Portmann}}, \bibinfo {author} {\bibfnamefont
  {D.}~\bibnamefont {Rosset}},\ and\ \bibinfo {author} {\bibfnamefont
  {M.}~\bibnamefont {Szegedy}},\ }\href
  {https://doi.org/10.1103/PhysRevLett.109.100401} {\bibfield  {journal}
  {\bibinfo  {journal} {Phys. Rev. Lett.}\ }\textbf {\bibinfo {volume} {109}},\
  \bibinfo {pages} {100401} (\bibinfo {year} {2012})}\BibitemShut {NoStop}%
\bibitem [{\citenamefont {Brassard}\ \emph {et~al.}(2016)\citenamefont
  {Brassard}, \citenamefont {Devroye},\ and\ \citenamefont
  {Gravel}}]{BrassardMulti}%
  \BibitemOpen
  \bibfield  {author} {\bibinfo {author} {\bibfnamefont {G.}~\bibnamefont
  {Brassard}}, \bibinfo {author} {\bibfnamefont {L.}~\bibnamefont {Devroye}},\
  and\ \bibinfo {author} {\bibfnamefont {C.}~\bibnamefont {Gravel}},\ }\href
  {https://doi.org/10.1109/TIT.2015.2504525} {\bibfield  {journal} {\bibinfo
  {journal} {IEEE Transactions on Information Theory}\ }\textbf {\bibinfo
  {volume} {62}},\ \bibinfo {pages} {876} (\bibinfo {year} {2016})}\BibitemShut
  {NoStop}%
\bibitem [{\citenamefont {Brassard}\ \emph {et~al.}(2019)\citenamefont
  {Brassard}, \citenamefont {Devroye},\ and\ \citenamefont
  {Gravel}}]{Brassard2}%
  \BibitemOpen
  \bibfield  {author} {\bibinfo {author} {\bibfnamefont {G.}~\bibnamefont
  {Brassard}}, \bibinfo {author} {\bibfnamefont {L.}~\bibnamefont {Devroye}},\
  and\ \bibinfo {author} {\bibfnamefont {C.}~\bibnamefont {Gravel}},\
  }\bibfield  {journal} {\bibinfo  {journal} {Entropy}\ }\textbf {\bibinfo
  {volume} {21}},\ \href {https://doi.org/10.3390/e21010092}
  {10.3390/e21010092} (\bibinfo {year} {2019})\BibitemShut {NoStop}%
\bibitem [{\citenamefont {Abramsky}\ and\ \citenamefont
  {Brandenburger}(2014)}]{abramsky2014operational}%
  \BibitemOpen
  \bibfield  {author} {\bibinfo {author} {\bibfnamefont {S.}~\bibnamefont
  {Abramsky}}\ and\ \bibinfo {author} {\bibfnamefont {A.}~\bibnamefont
  {Brandenburger}},\ }\bibinfo {title} {An {O}perational {I}nterpretation of
  {N}egative {P}robabilities and {N}o-{S}ignalling {M}odels, in {H}orizons of
  the {M}ind. {A} {T}ribute to {P}rakash {P}anangaden, edited by {F}. van
  {B}reugel, {E}. {K}ashefi, {C}. {P}alamidessi, and {J}. {R}utten}\ (\bibinfo
  {publisher} {Springer},\ \bibinfo {address} {Cham},\ \bibinfo {year} {2014})\
  pp.\ \bibinfo {pages} {59--75}\BibitemShut {NoStop}%
\bibitem [{\citenamefont {Pashayan}\ \emph {et~al.}(2015)\citenamefont
  {Pashayan}, \citenamefont {Wallman},\ and\ \citenamefont
  {Bartlett}}]{pashayan2015estimating}%
  \BibitemOpen
  \bibfield  {author} {\bibinfo {author} {\bibfnamefont {H.}~\bibnamefont
  {Pashayan}}, \bibinfo {author} {\bibfnamefont {J.~J.}\ \bibnamefont
  {Wallman}},\ and\ \bibinfo {author} {\bibfnamefont {S.~D.}\ \bibnamefont
  {Bartlett}},\ }\href {https://doi.org/10.1103/PhysRevLett.115.070501}
  {\bibfield  {journal} {\bibinfo  {journal} {Phys. Rev. Lett.}\ }\textbf
  {\bibinfo {volume} {115}},\ \bibinfo {pages} {070501} (\bibinfo {year}
  {2015})}\BibitemShut {NoStop}%
\bibitem [{\citenamefont {Onggadinata}\ \emph {et~al.}(2024)\citenamefont
  {Onggadinata}, \citenamefont {Kurzy{\'n}ski},\ and\ \citenamefont
  {Kaszlikowski}}]{us}%
  \BibitemOpen
  \bibfield  {author} {\bibinfo {author} {\bibfnamefont {K.}~\bibnamefont
  {Onggadinata}}, \bibinfo {author} {\bibfnamefont {P.}~\bibnamefont
  {Kurzy{\'n}ski}},\ and\ \bibinfo {author} {\bibfnamefont {D.}~\bibnamefont
  {Kaszlikowski}},\ }\href@noop {} {\bibfield  {journal} {\bibinfo  {journal}
  {arXiv}\ ,\ \bibinfo {pages} {(see today's submissions)}} (\bibinfo {year}
  {2024})}\BibitemShut {NoStop}%
\bibitem [{\citenamefont {Wigner}(1932)}]{wigner1932quantum}%
  \BibitemOpen
  \bibfield  {author} {\bibinfo {author} {\bibfnamefont {E.}~\bibnamefont
  {Wigner}},\ }\href {https://doi.org/10.1103/PhysRev.40.749} {\bibfield
  {journal} {\bibinfo  {journal} {Phys. Rev.}\ }\textbf {\bibinfo {volume}
  {40}},\ \bibinfo {pages} {749} (\bibinfo {year} {1932})}\BibitemShut
  {NoStop}%
\bibitem [{\citenamefont {Ferrie}\ and\ \citenamefont
  {Emerson}(2008)}]{ferrie2008frame}%
  \BibitemOpen
  \bibfield  {author} {\bibinfo {author} {\bibfnamefont {C.}~\bibnamefont
  {Ferrie}}\ and\ \bibinfo {author} {\bibfnamefont {J.}~\bibnamefont
  {Emerson}},\ }\href {https://doi.org/10.1088/1751-8113/41/35/352001}
  {\bibfield  {journal} {\bibinfo  {journal} {Journal of Physics A:
  Mathematical and Theoretical}\ }\textbf {\bibinfo {volume} {41}},\ \bibinfo
  {pages} {352001} (\bibinfo {year} {2008})}\BibitemShut {NoStop}%
\bibitem [{\citenamefont {Ferrie}\ and\ \citenamefont
  {Emerson}(2009)}]{ferrie2009framed}%
  \BibitemOpen
  \bibfield  {author} {\bibinfo {author} {\bibfnamefont {C.}~\bibnamefont
  {Ferrie}}\ and\ \bibinfo {author} {\bibfnamefont {J.}~\bibnamefont
  {Emerson}},\ }\href {https://doi.org/10.1088/1367-2630/11/6/063040}
  {\bibfield  {journal} {\bibinfo  {journal} {New Journal of Physics}\ }\textbf
  {\bibinfo {volume} {11}},\ \bibinfo {pages} {063040} (\bibinfo {year}
  {2009})}\BibitemShut {NoStop}%
\bibitem [{\citenamefont {Onggadinata}\ \emph
  {et~al.}(2023{\natexlab{a}})\citenamefont {Onggadinata}, \citenamefont
  {Kurzy\ifmmode~\acute{n}\else \'{n}\fi{}ski},\ and\ \citenamefont
  {Kaszlikowski}}]{onggadinata2023qubits}%
  \BibitemOpen
  \bibfield  {author} {\bibinfo {author} {\bibfnamefont {K.}~\bibnamefont
  {Onggadinata}}, \bibinfo {author} {\bibfnamefont {P.}~\bibnamefont
  {Kurzy\ifmmode~\acute{n}\else \'{n}\fi{}ski}},\ and\ \bibinfo {author}
  {\bibfnamefont {D.}~\bibnamefont {Kaszlikowski}},\ }\href
  {https://doi.org/10.1103/PhysRevA.107.032214} {\bibfield  {journal} {\bibinfo
   {journal} {Phys. Rev. A}\ }\textbf {\bibinfo {volume} {107}},\ \bibinfo
  {pages} {032214} (\bibinfo {year} {2023}{\natexlab{a}})}\BibitemShut
  {NoStop}%
\bibitem [{\citenamefont {Renes}\ \emph {et~al.}(2004)\citenamefont {Renes},
  \citenamefont {Blume-Kohout},\ and\ \citenamefont
  {Caves}}]{renes2004symmetric}%
  \BibitemOpen
  \bibfield  {author} {\bibinfo {author} {\bibfnamefont {J.~M.}\ \bibnamefont
  {Renes}}, \bibinfo {author} {\bibfnamefont {R.}~\bibnamefont
  {Blume-Kohout}},\ and\ \bibinfo {author} {\bibfnamefont {C.~M.}\ \bibnamefont
  {Caves}},\ }\href {https://doi.org/https://doi.org/10.1063/1.1737053}
  {\bibfield  {journal} {\bibinfo  {journal} {Journal of Mathematical Physics}\
  }\textbf {\bibinfo {volume} {45}},\ \bibinfo {pages} {2171} (\bibinfo {year}
  {2004})}\BibitemShut {NoStop}%
\bibitem [{\citenamefont {Kaszlikowski}\ and\ \citenamefont
  {Kurzy{\'n}ski}(2021)}]{kaszlikowski2021little}%
  \BibitemOpen
  \bibfield  {author} {\bibinfo {author} {\bibfnamefont {D.}~\bibnamefont
  {Kaszlikowski}}\ and\ \bibinfo {author} {\bibfnamefont {P.}~\bibnamefont
  {Kurzy{\'n}ski}},\ }\href
  {https://doi.org/https://doi.org/10.1007/s10701-021-00461-w} {\bibfield
  {journal} {\bibinfo  {journal} {Foundations of Physics}\ }\textbf {\bibinfo
  {volume} {51}},\ \bibinfo {pages} {55} (\bibinfo {year} {2021})}\BibitemShut
  {NoStop}%
\bibitem [{\citenamefont {Onggadinata}\ \emph
  {et~al.}(2023{\natexlab{b}})\citenamefont {Onggadinata}, \citenamefont
  {Kurzynski},\ and\ \citenamefont
  {Kaszlikowski}}]{onggadinata2023simulations}%
  \BibitemOpen
  \bibfield  {author} {\bibinfo {author} {\bibfnamefont {K.}~\bibnamefont
  {Onggadinata}}, \bibinfo {author} {\bibfnamefont {P.}~\bibnamefont
  {Kurzynski}},\ and\ \bibinfo {author} {\bibfnamefont {D.}~\bibnamefont
  {Kaszlikowski}},\ }\href {https://doi.org/10.1103/PhysRevA.108.032204}
  {\bibfield  {journal} {\bibinfo  {journal} {Phys. Rev. A}\ }\textbf {\bibinfo
  {volume} {108}},\ \bibinfo {pages} {032204} (\bibinfo {year}
  {2023}{\natexlab{b}})}\BibitemShut {NoStop}%
\bibitem [{\citenamefont {Mermin}(1990)}]{mermin1990extreme}%
  \BibitemOpen
  \bibfield  {author} {\bibinfo {author} {\bibfnamefont {N.~D.}\ \bibnamefont
  {Mermin}},\ }\href {https://doi.org/10.1103/PhysRevLett.65.1838} {\bibfield
  {journal} {\bibinfo  {journal} {Phys. Rev. Lett.}\ }\textbf {\bibinfo
  {volume} {65}},\ \bibinfo {pages} {1838} (\bibinfo {year}
  {1990})}\BibitemShut {NoStop}%
\end{thebibliography}%


\end{document}